\documentclass[prl,twocolumn,superscriptaddress,amsmath,amsfonts,amssymb,preprintnumbers]{revtex4}
\usepackage{graphicx}
\usepackage[colorlinks,plainpages=false,linkcolor=black,urlcolor=blue,citecolor=black,pdfpagemode=UseNone,pdfstartview=FitBH]{hyperref}
\usepackage{sidecap}

\makeatletter
\renewcommand{\@evenfoot}{\hfill\raisebox{-3em}{\bf\thepage}\hfill}
\renewcommand{\@oddfoot}{\hfill\raisebox{-3em}{\bf\thepage}\hfill}
\makeatother

\begin{document}

\title{Giant Nernst effect and bipolarity in the quasi-one-dimensional metal, Li$_{0.9}$Mo$_6$O$_{17}$}

\author{J.~L.~Cohn}
\affiliation{Department of Physics, University of Miami, Coral Gables, FL 33124}

\author{B.~D.~White}
\affiliation{Department of Physics, Montana State University, Bozeman, Montana 59717}

\author{C.~A.~M. dos Santos}
\affiliation{Escola de Engenharia de Lorena - USP, P. O. Box 116, Lorena-SP, 12602-810, Brazil}

\author{J.~J.~Neumeier}
\affiliation{Department of Physics, Montana State University, Bozeman, Montana 59717}

\begin{abstract}
The Nernst coefficient for the quasi-one-dimensional metal, Li$_{0.9}$Mo$_6$O$_{17}$,
is found to be among the largest known for metals ($\nu\simeq 500\ \mu$V/KT at
$T\sim 20$~K), and is enhanced in a broad range of temperature by orders of magnitude over the value expected from
Boltzmann theory for carrier diffusion.  A comparatively small Seebeck coefficient implies that Li$_{0.9}$Mo$_6$O$_{17}$
is bipolar with large, partial Seebeck coefficients of opposite sign.  A very large thermomagnetic figure of merit, $ZT\sim 0.5$,
is found at high field in the range $T\approx 35-50$~K.
\end{abstract}

\maketitle
\thispagestyle{empty}\clearpage

The Nernst effect, a transverse electric field generated by a longitudinal temperature gradient in the
presence of a perpendicular magnetic field, has been known since its discovery in the 19th century
in the semimetal bismuth \cite{Ettingshausen}. Conductors with large Nernst coefficients are potential candidates
for use in cryogenic refrigeration \cite{FreibertReview}, but in spite of several decades' research and
a demand for such devices at temperatures below 100 K (e.g., for infrared detectors),
none have yet proved practical.
In recent years, Nernst-effect measurements have received renewed attention as a means for probing the pseudogap phase of
cuprate superconductors \cite{WangReview,Taillefer} and novel ground states of other correlated-electron systems \cite{BehniaReview},
prompting a reassessment of the physics leading to large Nernst coefficients in metals.
Emphasizing the contribution from diffusing charge carriers, experiment and theory agree \cite{BehniaReview} that
the Nernst-coefficient magnitude is set by the ratio of charge carrier mobility to Fermi energy.
Here we report observations of very large Nernst coefficients
in the quasi-one-dimensional (Q1D) metal, Li$_{0.9}$Mo$_6$O$_{17}$, rivaling the largest known for bismuth and graphite.
This material possesses \emph{neither} a very large mobility nor a small Fermi energy.  We establish that
conduction along the Q1D chains of this material is bipolar, with very large partial thermopowers of opposite sign that predominate
over conventional carrier diffusion in a broad temperature range.  The observations offer new insight into the origin of large
thermoelectric coefficients in bulk conductors and expand the criteria for viable
materials in thermomagnetic cooling applications.

Li$_{0.9}$Mo$_6$O$_{17}$, a low-temperature superconductor ($T_c\approx 2$~K) first synthesized and studied in the
1980s \cite{PBReview,oldwork2}, has attracted interest more recently for its quasi-one dimensionality and Luttinger-liquid
candidacy \cite{PES2,BandStructure1,PES3,Tunneling}.  This compound is unusual among inorganic and organic Q1D compounds in its
absence of a conventional density-wave transition \cite{Optical,NeumeierPRL} (either charge or spin) throughout a broad temperature range,
$T\geq T_c$. An upturn in its resistivity below $T_M\sim 30$~K may be associated with localization, dimensional crossover or the development of unconventional
(e.g., electronically-driven) charge density-wave order \cite{Optical,NeumeierPRL,XuPRL}.
Though the $T\to 0$ ground state (superconducting, metallic or insulating) is sensitive to heat treatment and can vary among as-prepared crystals,\cite{Neumeier2band}
the anisotropy and temperature variation of electrical properties for $T\geq 5$~K, the regime of study here, are robust and
reproducible \cite{Optical,NeumeierPRL,XuPRL,Neumeier2band,NeumeierMontgomery,ChineseMR}.

Single-crystal growth of Li$_{0.9}$Mo$_6$O$_{17}$ is described in detail elsewhere \cite{NeumeierPRL,oldwork2}.
Crystals had typical dimensions $0.05-0.15$~mm along $a$ and $0.5-2$~mm in the $b$-$c$ plane.  Longitudinal and transverse
voltages were measured with Au leads attached with silver epoxy.  A 25~$\mu$m-diameter chromel-constantan differential thermocouple
monitored the temperature difference.  The Nernst signal at each field was determined from the slope of linear $V_H$-$\Delta T_x$ curves
(with $V_H$ the field-odd transverse voltage). Lattice constants and oxygen content for six crystals studied
and additional details on the transport measurements are described in the Supplemental Material \cite{SMO}.
\begin{figure*}
\includegraphics[width=6.4in]{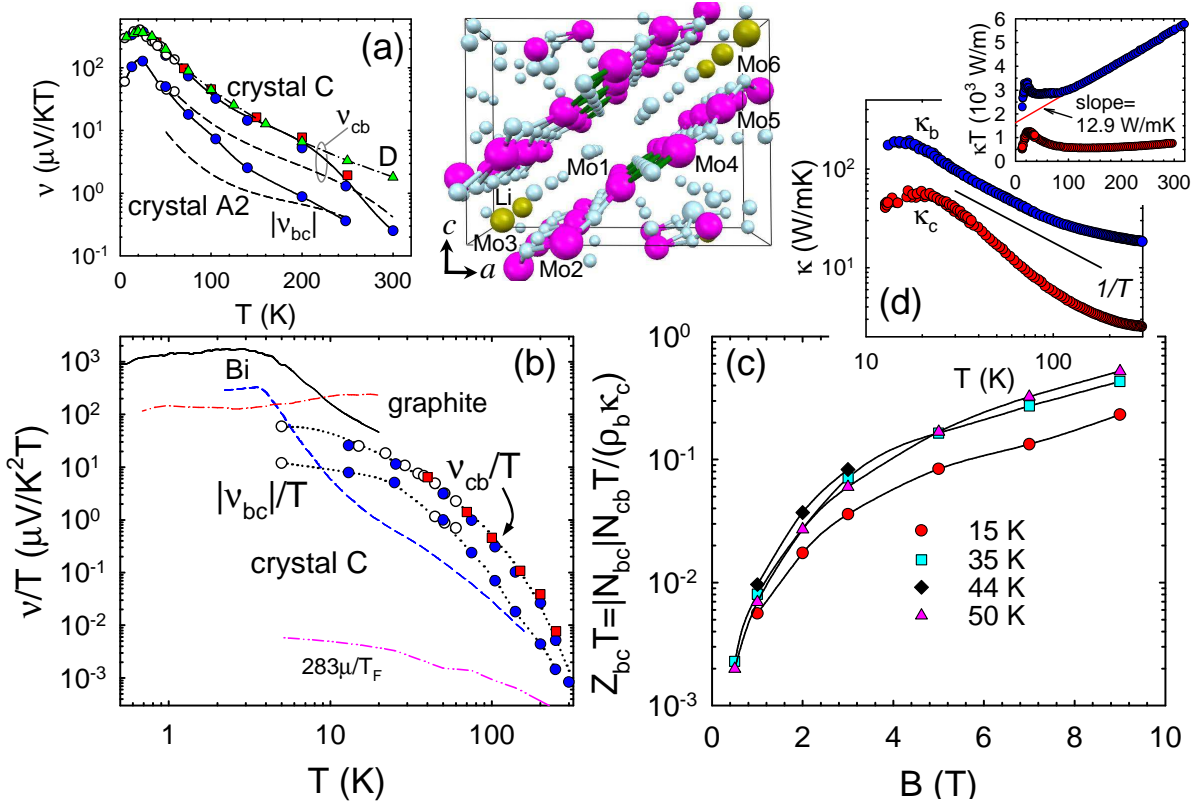}
\caption{(color online) (a) Nernst coefficient, $\nu(T)$, for crystals A2 (dashed lines), C (circles and squares), and D (triangles). For crystal C: $\nu_{cb}$ and $\nu_{bc}$
determined with primary heat current along $b$ (solid circles) and $c$ (solid squares), respectively.
The specimen, originally with thickness $110\ \mu$m along $a$, was mechanically thinned to $55\ \mu$m and remeasured with primary heat current
along $b$ (open circles) \cite{SMO}. (b) $\nu/T$ {\it vs.} $T$ for crystal C, compared to bismuth
[Ref.'s~\onlinecite{BehniaBi} (solid curve) and \onlinecite{Korenblit} (dashed curve)] and graphite (Ref.~\onlinecite{BehniaGraphite}).
The dash-dot-dot curve labeled ``$283\mu/T_F$'' represents the carrier-diffusion contribution from
Boltzmann theory, determined from the carrier mobilities and estimated Fermi temperature \cite{SMO}.
(c) Figure of merit as a function of magnetic field for crystal C at several
temperatures. (d) zero-field thermal conductivities along the \emph{b} and \emph{c} axes, inset: same data plotted
as $\kappa T$ {\it vs.} $T$ to emphasize the large constant term in $\kappa_b(T)$.
Main Inset: Perspective view of the LMO monoclinic unit cell viewed along the \emph{b} axis; Q1D Mo1-O-Mo4 zigzag double chains composed of MoO$_6$ octahedra
(highlighted bonds) are part of planar Mo-O networks whose two dimensionality is broken by the Li ions.}
\label{NernstvsT}
\end{figure*}

Complete sets of transport measurements (resistivity, thermopower, thermal conductivity, Hall and Nernst coefficients) were measured on two single crystals
of Li$_{0.9}$Mo$_6$O$_{17}$ (LMO), designated A2 and C, with electric or heat currents along the Q1D Mo-O chains (crystallographic \emph{b} axis; inset Fig. 1).
The Nernst coefficient of specimen C was also measured with heat current along the \emph{c} axis.  Thermal and thermoelectric measurements
(heat flow along \emph{b}) were also performed on a third crystal, designated D. For heat flow along the $x$ direction (temperature gradient, $\nabla_x T$) and
magnetic field along $z$, the Nernst coefficient is defined as, $\nu_{yx}=-E_y/(\nabla_x TB_z).$

With magnetic field along the least conducting \emph{a} axis, $\nu_{cb}$ and $\nu_{bc}$ were determined for specimens A2 and C and $\nu_{cb}$ for specimen D as
shown in Fig.\ref{NernstvsT}(a) \cite{SMO}.  The observed ratio, $\nu_{cb}/|\nu_{bc}|\sim 3-6$, reflects the
resistivity anisotropy, $\rho_c/\rho_b$, and agrees well with measurements on similar crystals \cite{NeumeierMontgomery,ChineseMR}.
$\nu(T)$ in nonmagnetic metals may have contributions from charge-carrier diffusion and from phonon drag, the latter arising from the
interaction of the carriers with a non-equilibrium population of heat-carrying phonons.  The
diffusion term can be expressed as \cite{Ussishkin,BehniaReview}, $\nu/T\simeq (\pi^2/3)(k_B/e)(\mu/T_F)=283(\mu/T_F)$,
where $\mu$ is the carrier mobility and $T_F$ the Fermi temperature.  This expression approximates $\nu/T$ quite well at low $T$ (within a factor of $3-5$) for
Bi, graphite and a variety of other multi-band, correlated, and/or low-dimensional metals \cite{BehniaReview}, over six orders of magnitude in
$\mu/T_F$ \cite{SMO}.  The $\nu/T$ data for LMO [Fig.~\ref{NernstvsT}(b)] exceed an estimate of this diffusion contribution
(dash-dot-dot curve) by four orders of magnitude at low $T$, and remain significantly larger
to room temperature.  The expression is strictly valid for a linear energy dependence of the Hall angle at the Fermi level.
Though strong energy-dependent scattering, as may arise from electron correlations or incipient
density-wave order, or proximity to a quantum critical point\cite{BehniaReview}, can enhance this energy dependence
and hence $\nu/T$, it is very unlikely such effects can account
for this very large discrepancy \cite{Ussishkin}.  We conclude that a mechanism other than carrier diffusion predominates
in $\nu(T)$ over the entire temperature range.

Consider the potential of LMO for application.
For operation as an Ettingshausen cooler, with electric current applied along the \emph{b} axis generating heat flow along \emph{c}, the
relevant adiabatic thermomagnetic figure of merit is \cite{FreibertReview},
$Z_{bc}T=N_{bc}N_{cb}T/(\rho_b\kappa_c)$ where $N=\nu B$ and $\kappa_c$ is the \emph{c}-axis thermal conductivity. Fig.~\ref{NernstvsT}(c)
shows the field dependence of $Z_{bc}T$ along with $\kappa_c(T)$ [Fig.~\ref{NernstvsT}(d)]; the latter is independent of field.  We see that $ZT$ at 35~K
reaches 0.5 for B=9T with no sign of saturation,
among the highest known values for any material \cite{BiSbZT,BehniaPFP}.  Viable thermomagnetic materials for applications should
have $ZT$ approaching 1 at fields achievable using small permanent magnets ($B\lesssim 1$T). It is conceivable that
the inter-chain $\kappa_c$ can be reduced (e.g. through the introduction of mass disorder via chemical substitution)
to enhance $ZT$ at lower field.
\begin{figure}[t]
\includegraphics[width=\columnwidth]{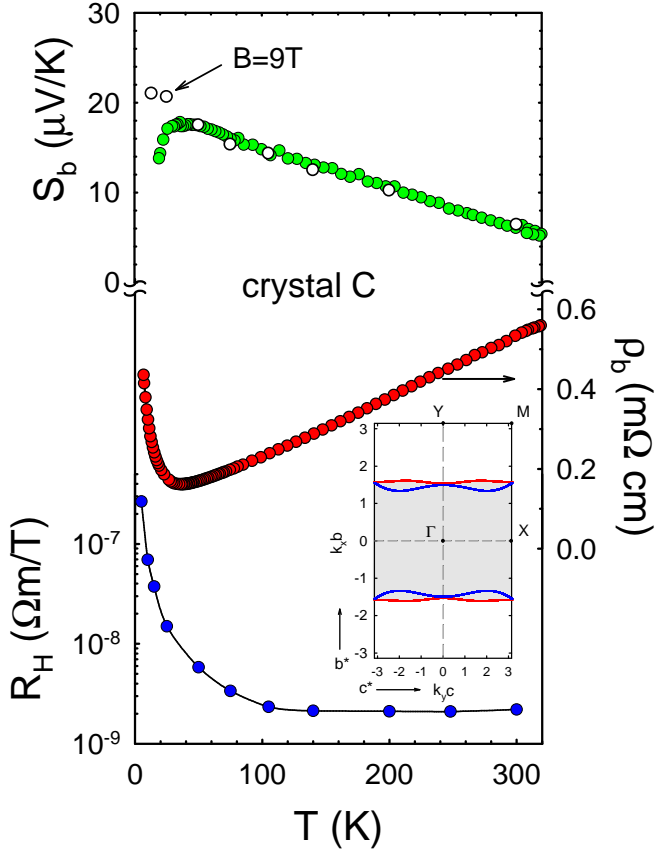}
\caption{(color online) Seebeck coefficient ($S$), resistivity, and
Hall coefficient for specimen C. Inset: Fermi surface for Li$_{0.9}$Mo$_6$O$_{17}$ (adapted from Ref.~\onlinecite{BandStructure1}).}
\label{crystalC}
\end{figure}
\begin{figure}[t]
\includegraphics[width=\columnwidth]{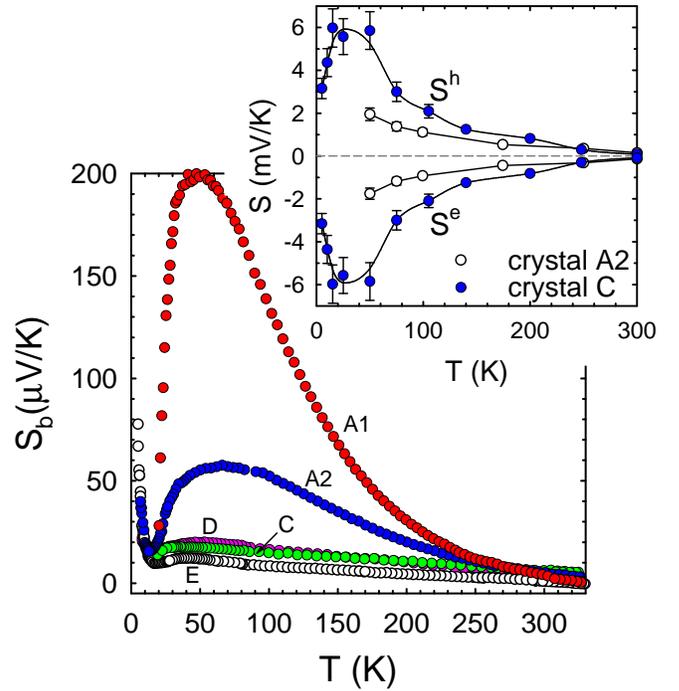}
\caption{(color online) Zero-field {\it b}-axis Seebeck coefficient for four Li$_{0.9}$Mo$_6$O$_{17}$ specimens.
Combined results of x-ray, compositional, and Hall measurements indicate that specimens with large peaks (A1, A2) are more oxygenated
and particle-hole asymmetric ($R^h\gg R^e$ and $\sigma^h<\sigma^e$) compared to those with linear-$T$ thermopowers (C,D,E) \cite{SMO}.
Inset: Hole and electron partial thermopowers computed using the equation \cite{Eqn}, the measured Nernst coefficient ($\nu_{cb}$),
and two-band model parameters; error bars (25\%) are propagated from those of the Hall data \cite{SMO}.}
\label{TEP}
\end{figure}

The Seebeck coefficient or thermopower (TEP), $S=-E_x/\nabla_x T$, reveals an important aspect of conduction in LMO.
Crystals C and D with the largest $\nu(T)$ have TEP values (Fig.'s~\ref{crystalC} and \ref{TEP}) that are modest in magnitude.
Consequently, the ratio of transverse to longitudinal thermoelectric fields, $E_y/E_x=\nu B/S$,
exceeds $10^2$ at 40~K; the electric field is essentially perpendicular to the
temperature gradient. The implication is that LMO is compensated, with a cancellation of very large hole and electron
Seebeck coefficients (opposite in sign).
This can be seen more explicitly from the expression for $\nu$ in a bipolar, anisotropic
conductor \cite{Eqn} which includes a bipolar term involving the difference, $S^h-S^e$, of the hole and
electron partial Seebeck coefficients along the heat flow.  For perfect compensation,
$\nu_{cb}\simeq (1/2)\mu_b\left(S_b^h-S_b^e\right)$, while $S_b=(1/2)(S_b^h+S_b^e)$.  Bipolarity is plausible given the band
structure \cite{PES2,BandStructure1,BandStructure2} composed of two nearly-degenerate, approximately 1/2-filled, Q1D bands
of Mo $d_{xy}$ parentage crossing the Fermi energy; there are two pairs of slightly warped Fermi surface (FS) sheets in the
\emph{b*-c*} planes (inset, Fig.~\ref{crystalC}).  Each sheet, with positive and negative curvature, can contribute electron and
hole character if the mean free path is anisotropic \cite{OngGeometricHall}.  Bipolarity is well known to enhance the Nernst effect and is a
prerequisite for useful thermomagnetic materials.

Supporting this bipolar picture, the TEP of other crystals with higher oxygen content,
like A1 and A2 (Fig.~\ref{TEP}), exhibit very large, positive maxima at $T=50-70$~K
that augment the linear-$T$ form (specimens C, D, E) up to 300~K and beyond (previously published data for the TEP of LMO \cite{Boujida}
fall between those of specimens A2 and C). This behavior and the smaller Nernst coefficient for specimen A2
are consistent with a departure from electron-hole symmetry in the more oxygenated crystals and incomplete cancellation of very large
partial thermopowers.

A two-component analysis of the transport places these qualitative observations on a quantitative footing.
Magnetoresistance and Hall data \cite{SMO} for
crystal C yield $R^h\simeq -R^e\simeq 3.4\times 10^{-8}\ \Omega$~m/T at 300 K and $\sigma^h\simeq \sigma^e$ such that the measured
Hall coefficient (Fig.~\ref{crystalC}) is an order of magnitude smaller than $R^h$ or $|R^e|$, consistent
with the substantial cancellation implied by the thermoelectric coefficients.  The effective carrier density
$\approx 2/(R^h|e|)\simeq 3.7\times 10^{26}$~m$^{-3}$ represents $\sim 13\%$ of that estimated from the
chemistry/bonding \cite{SMO}, suggesting localization of a significant
fraction of carriers at the FS in accord with optical studies \cite{Optical}.  A similar analysis of transport data for
crystal A2 \cite{SMO} yields $R^h\simeq 7|R^e|$, $\sigma^h\simeq 0.4\sigma^e$, and a mobile carrier fraction $\sim 2.5\%$.
Using these hole and electron transport coefficients and the measured $\nu_{cb}(T)$, the equation for $\nu$ \cite{Eqn} can be inverted to compute $S_b^h(T)$ and $S_b^e(T)$ for the
two crystals (inset, Fig.~\ref{TEP}).  In generating these curves it was assumed that $\sigma^h_c/\sigma_c=\sigma^h_b/\sigma_b$, but relaxing this assumption
yields comparable TEP magnitudes \cite{SMO}.  The colossal TEP values ($2-6$~mV/K at low $T$) are
typical of semiconductors \cite{FeSb2}, but are unprecedented for metals.

We hypothesize that the computed TEP is substantially
larger than the intrinsic chain-axis value because the measured $\sigma_b$
in LMO is suppressed by the occurrence of incomplete electrical connectivity of the Mo-O chains.
This hypothesis is motivated by the Hall data implying a large fraction of
localized carriers, and the observation of a large constant term in $\kappa_b(T)$ ($12.9$~W/mK) at $T\geq 80$~K [inset,
Fig.~\ref{NernstvsT}(d)]. The latter is attributed to electronic heat conduction.  Poor connectivity will substantially suppress the
measured $\sigma_b$ from that of continuous chains, whereas electronic heat conduction on chain fragments will continue to
contribute given that heat is transferred readily to the lattice. Supporting this picture, the constant term in $\kappa_b$ varies
little among several other crystals with varying $\sigma_b$ \cite{SMO}.
This scenario is appealing because it can simultaneously account for the large fraction of total charge that is localized and the
different values for the anisotropy ratio $\sigma_b/\sigma_c$ reported
in the literature \cite{PBReview,NeumeierMontgomery,ChineseMR,Wakeham}.  It also
favors a Sommerfeld value for the Lorenz number ($L_0=2.45\times 10^{-8}\ {\rm W\Omega/K^2}$): the resistivity for specimen C and
its estimated mobile carrier fraction ($f=0.13$) from the Hall measurement yields, $\kappa_b^e(300~{\rm K})=(L_0/f)\times
0.55\times 10^{-5}(\Omega {\rm m})^{-1}\times 300$~K=10~W/mK.  A bipolar thermal conductivity \cite{bipolarK}, expected
in a metal with overlapping bands at $E_F$, should also contribute $\lesssim 1$~W/mK.  Bipolar heat flow
should dominate the thermal Hall conductivity (Righi-Leduc effect) of a closely compensated system, offering an alternative explanation for
recent observations of an apparent Wiedemann-Franz law violation \cite{Wakeham} in LMO based on such measurements.
An intrinsic chain resistivity at 300~K, $\rho_{ch}=f\rho_b\simeq 70\ \mu\Omega$cm and Hall mobility
$\mu_{ch}\simeq (\mu_b/f)\simeq 2.3\times 10^{-2}{\rm T}^{-1}$ are inferred. The computed partial thermopowers (Fig.~\ref{TEP}),
depending inversely on $\sigma_b$ \cite{Eqn}, should be reduced accordingly by the factor $f$.

On the other hand, the measured thermoelectric coefficients reflect appropriate bulk averages, weighing voltage contributions
from short and long chain fragments according to local temperature differences.  Thus the intrinsic chain thermoelectric behavior must be closest to
that of the most conducting specimen C, having the largest value of $\nu$ and smallest of the TEP (the most closely compensated).
The deviation from particle-hole symmetry and increased carrier localization
characterizing more oxygenated crystals presumably reflect the combined effects of charge doping and disorder associated with oxygen defects \cite{Neumeier2band}.
The electronic system of LMO is most one-dimensional at high $T$ where the thermal energy exceeds the transverse hopping energy ($t_{\perp}$) that determines the
warping of the FS sheets \cite{Giamarchi}.  This is the regime where a conventional phonon-drag mechanism is least likely to explain the large thermoelectric coefficients since
anharmonic phonon-phonon scattering should limit momentum transfer to the electron system, rendering phonon drag negligible.  However, it is possible
the low dimensionality of the lattice \cite{NeumeierPRL} suppresses phonon-phonon scattering along the chains \cite{Morelli}, thereby enhancing phonon drag in this regime.
It is difficult to assess the possible relevance of Luttinger physics.  Theoretical treatments of the thermopower
for single Luttinger chains with impurity scattering \cite{LuttingerTEP}
indicate a linear temperature dependence, inconsistent with the present results.
More realistic models including electron-electron Umklapp scattering and coupled chains (essential for the Nernst effect)
have not been treated to our knowledge.
At $T\leq T_M$, where higher-dimensional behavior is manifested in unconventional
density-wave order \cite{Optical,NeumeierPRL,XuPRL} and the appearance of superconductivity, large phonon-drag effects are plausible, particularly near $T=20$~K where both $\kappa_b$ and $\nu$
exhibit maxima (Fig.~\ref{NernstvsT}). There the phonon mean free path $\Lambda$ is $\sim 10\ \mu$m, as inferred from kinetic theory using
the measured specific heat \cite{Optical} and an acoustic phonon velocity $\upsilon\simeq 3$~km/s.  This implies a phonon-drag TEP \cite{2DEGs},
$S_g\simeq \upsilon\Lambda/\mu_{e-ph}T\approx 750\ \mu$V/K using $\mu_{e-ph}=2{\rm T}^{-1}$ (estimated as 5 times the average $\mu/f$ for specimen C),
in good agreement with the computed partial TEP value (corrected by the factor $f$).

In summary, Li$_{0.9}$Mo$_6$O$_{17}$ is found to have a bipolar Nernst coefficient that is among the largest known for metals and far exceeds
the expected contribution from carrier diffusion over a broad temperature range.
Though phonon drag appears capable of accounting for this discrepancy at the lowest $T$, low-dimensional physics of the electronic and/or lattice
systems may be important to a complete understanding of its enhanced thermoelectric coefficients.

The authors acknowledge very helpful input from A. L. Chernyshev.  This material is based upon work supported by the National Science
Foundation under grant DMR-0907036 (Mont.~St.~Univ.), the Research Corporation (Univ.~Miami), and
in Lorena by the CNPq (301334/2007-2) and FAPESP (2009/14524-6).

\newpage
\setcounter{figure}{0}
\renewcommand{\figurename}{Fig. S\!}
\onecolumngrid
\newpage
\begin{center}
{\large\bf Supplementary Material}
\vskip0.17in
J.~L.~Cohn, B.~D.~White, C.~A.~M. dos Santos, and J.~J.~Neumeier
\vskip0.05in
\end{center}

\subsection*{Properties of the crystals studied}
\vspace{-.25in}
\begin{table}[h]
\caption{Lattice constants, oxygen content as determined by energy-dispersive x-ray spectroscopy, and
Hall coefficient (at 250~K) for the six Li$_{0.9}$Mo$_6$O$_{17}$
crystals studied in this paper.}
\begin{ruledtabular}
\begin{tabular}{cccccccc}
Specimen &$a$ (\AA)&$b$ (\AA)&$c$ (\AA)
 &Wt.\% O &$R_H\ (10^{-9}\Omega$~m/T)\\
A1& 12.751(2) & 5.521(2) & 9.475(5) &32.1(3)&-- \\
A2\footnotemark[1] & 12.752(2) & 5.520(2) & 9.482(4) &--&3.3 \\
B & 12.751(2) & 5.520(2) & 9.486(5) &30.6(4)&-- \\
C& 12.753(2) & 5.520(2) & 9.491(4) &--&2.1 \\
D& 12.753(2) & 5.519(2) & 9.488(4) &--& --\\
E& 12.751(2) & 5.520(2) & 9.489(3) &29.7(3)&-- \\
\end{tabular}
\end{ruledtabular}
\footnotetext[1]{crystal A2 is the same as specimen A1, but measured 18 months later with fresh electrical contacts.
The implication is that crystals tend to lose oxygen over time.
}
\end{table}
\vspace{-.8in}
\begin{SCfigure}[][b]
\includegraphics[width=3.4in,clip]{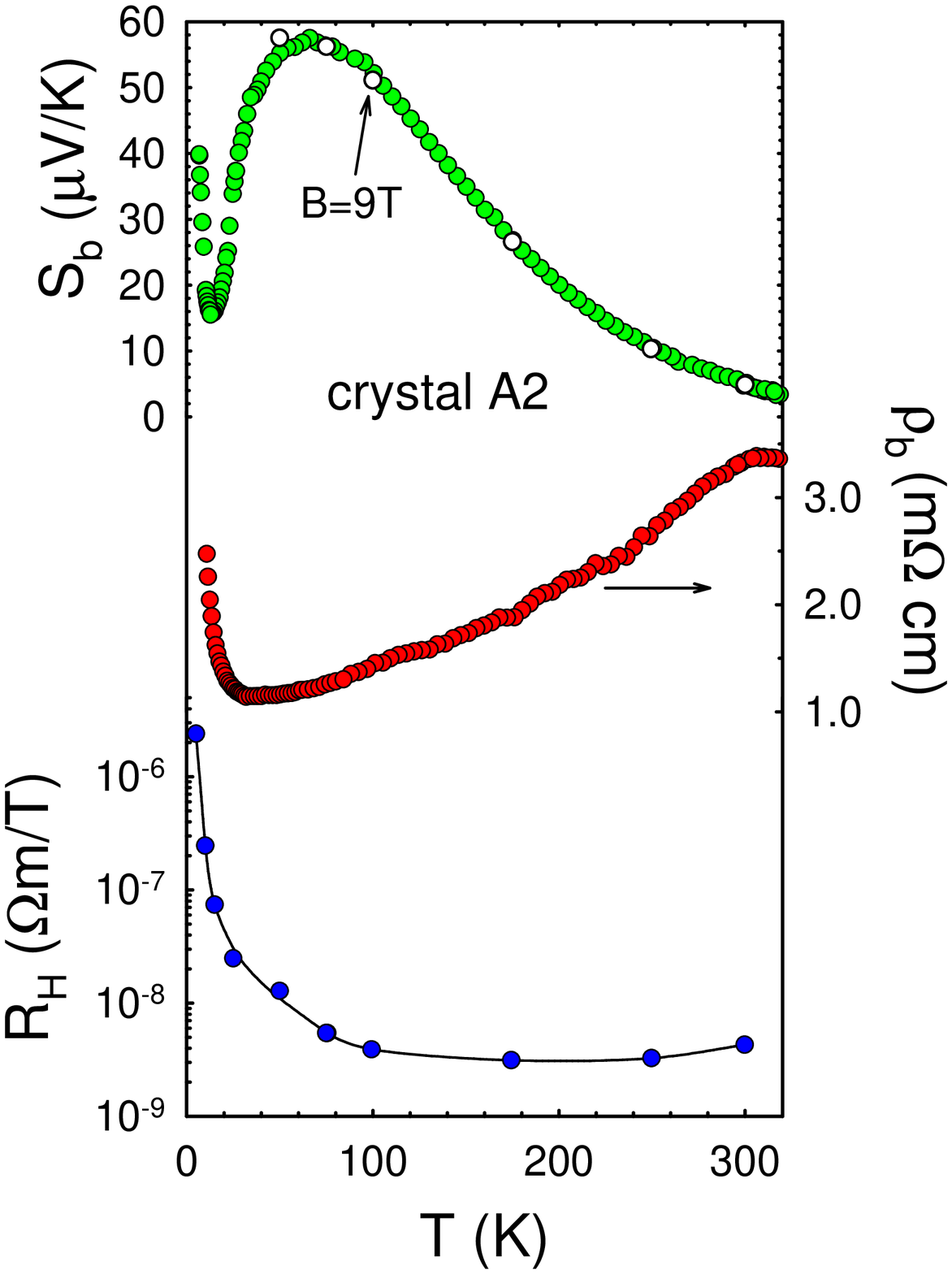}
\vspace{-.8in}
\caption{Seebeck coefficient ($S$), resistivity,
and Hall coefficient for specimen A2 along the conducting chains.}
\label{S1}
\end{SCfigure}
\vspace{0.3in}
\section*{Transport measurements}
For thermoelectric measurements specimens were suspended in vacuum with one end attached to a Cu cold sink (silver epoxy)
and the other affixed with a small metal-film chip heater (varnish).  Longitudinal and transverse
voltages were measured with Au leads attached with silver epoxy.  A 25~$\mu$m-diameter chromel-constantan differential thermocouple
monitored the temperature difference across the sample and a separate chromel-constantan thermocouple monitored the average
sample temperature (relative to the calibrated thermometer mounted in the cold sink).  A high-precision, differential method was employed at fixed magnetic fields to
determine the Nernst coefficient: at each field the heater current was cycled through three
different values and voltages measured after a steady state was achieved.  This process was repeated for reversed magnetic field.
The Nernst signal at each field was determined from the slope of resulting linear $V_H$-$\Delta T_x$ curves, with
$V_H=[V_{xy}(B)-V_{xy}(-B)]/2$ the field-odd transverse voltage (Fig.~S2~a). The Nernst signal was linear in magnetic field at all temperatures
for $B\lesssim 3$~T (Fig.~S2~b).  Nonlinearity at higher fields was evident at temperatures $T\lesssim 35$~K.  The Nernst coefficient
was computed from the low-field slope in these plots. The field dependence of the thermocouple\cite{Inyushkin} for
temperatures $T\geq 10$~K is $\lesssim 5$\%.

The Hall coefficient was measured during the same experimental runs.  Linearity of the Hall voltage in applied current and magnetic field
was observed for $T\geq 50$~K.  Some nonlinearity in field was observed at lower temperatures.
The Hall coefficients of both specimens A2 and C were re-measured in subsequent experiments under isothermal
conditions (varnished flat to the heat sink); good agreement was found with both sets of measurements.

For thermal-conductivity measurements, heat losses associated with radiation and conduction through the leads were
assessed in separate experiments with specimens suspended from their leads in vacuum.  The data presented in Fig.~1~d
have been corrected accordingly.  These corrections amounted to 30\% near 300~K for $\kappa_c$ and 5\% for $\kappa_b$.  At $T\leq 100$~K
the corresponding corrections were $<6$\% and $<1$\%, respectively.

\begin{figure*}[h]
\includegraphics[width=5.5in,clip]{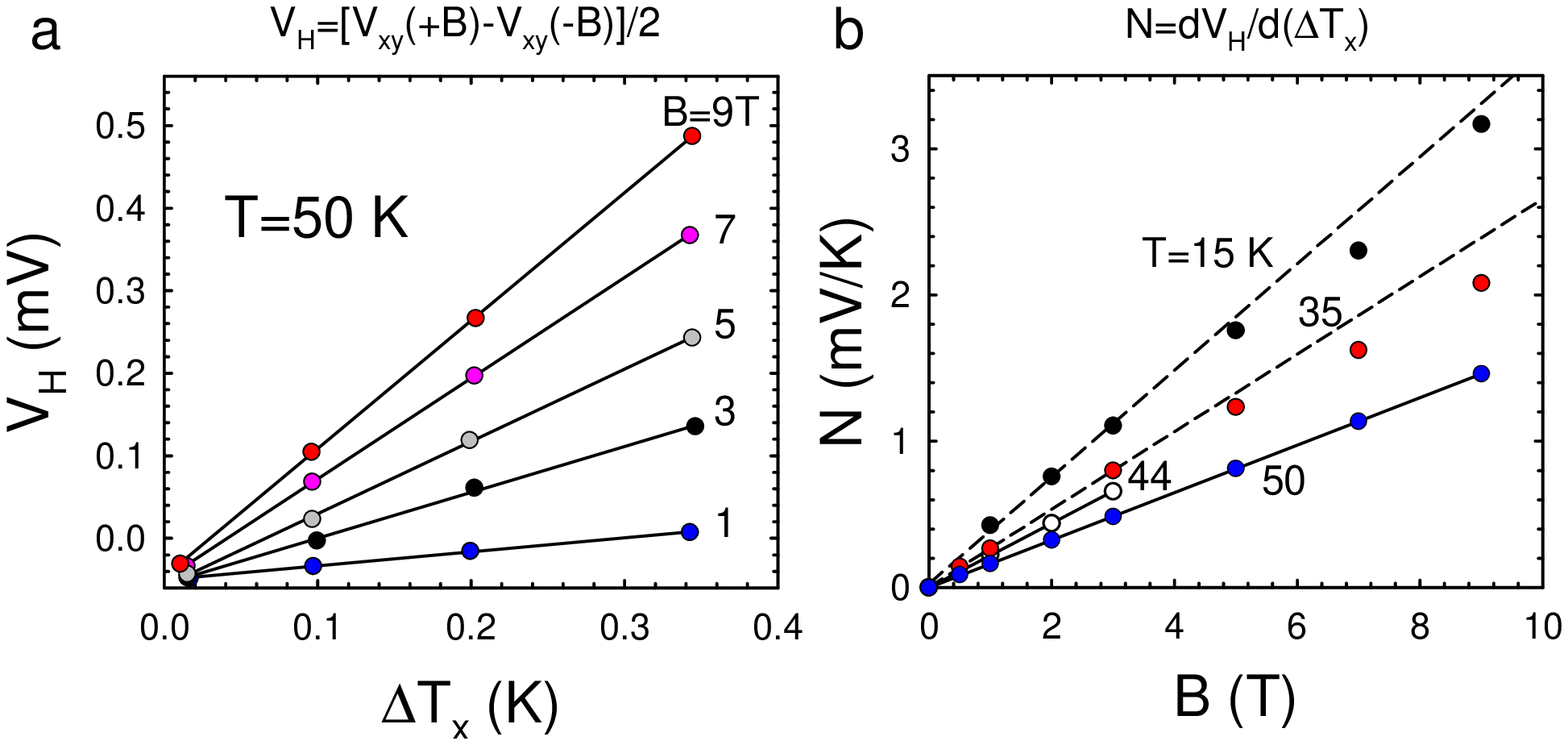}
\caption{Variation of the Nernst voltage with longitudinal
temperature difference at fixed fields ({\bf a}) and field dependence of the Nernst signal
at fixed temperatures ({\bf b}).} \label{S2}
\end{figure*}

The heat current in highly anisotropic crystals always has a small component along the
nominally transverse direction during Nernst measurements. This induces a field-odd Nernst voltage
across the \emph{longitudinal} voltage leads. Thus, for an experiment with heat flow along $b$, $\nu_{bc}$ can be determined
by extracting both the field-odd and field-even
parts of $V_{xx}$ and $\Delta T_y$, respectively, the latter determined with an additional transverse thermocouple.
This procedure yielded excellent agreement between values for $\nu_{cb}$
extracted from separate experiments on specimen C (discussed in Fig.~1) with heat currents
$J_q||c$ and $J_q||b$.  A field-odd component of $\Delta T_y$ produced by the Righi-Leduc effect
was small, typically $<10^{-2}\Delta T_x$, consistent with the small Hall angle.

\section*{Two-carrier analysis of magnetotransport}

The magnetoresistance (MR) and Hall resistivity of a conductor with two
carrier types and field-independent densities and mobilities can
be expressed as\cite{Pippard,Chambers},
$$\Delta\rho(B)/\rho={MR_{\infty}(\mu_{eff}B)^2\over 1+(\mu_{eff}B)^2}\qquad {\rm and}\qquad \rho_H(B)={R_{H0}+R_{H\infty}(\mu_{eff}B)^2\over
1+(\mu_{eff}B)^2}B$$
\noindent
where,
$$MR_{\infty}={(\sigma^hR^h-\sigma^eR^e)^2\over \sigma^h\sigma^e(R^h+R^e)^2},\qquad \qquad \mu_{eff}={\sigma^h\sigma^e(R^h+R^e)\over \sigma^h+\sigma^e}$$
$$R_{H0}={(\sigma^h)^2R^h+(\sigma^e)^2R^e\over (\sigma^h+\sigma^e)^2},\qquad \qquad R_{H\infty}={R^hR^e\over R^h+R^e}$$

\begin{figure}[h]
\includegraphics[width=3.5in,clip]{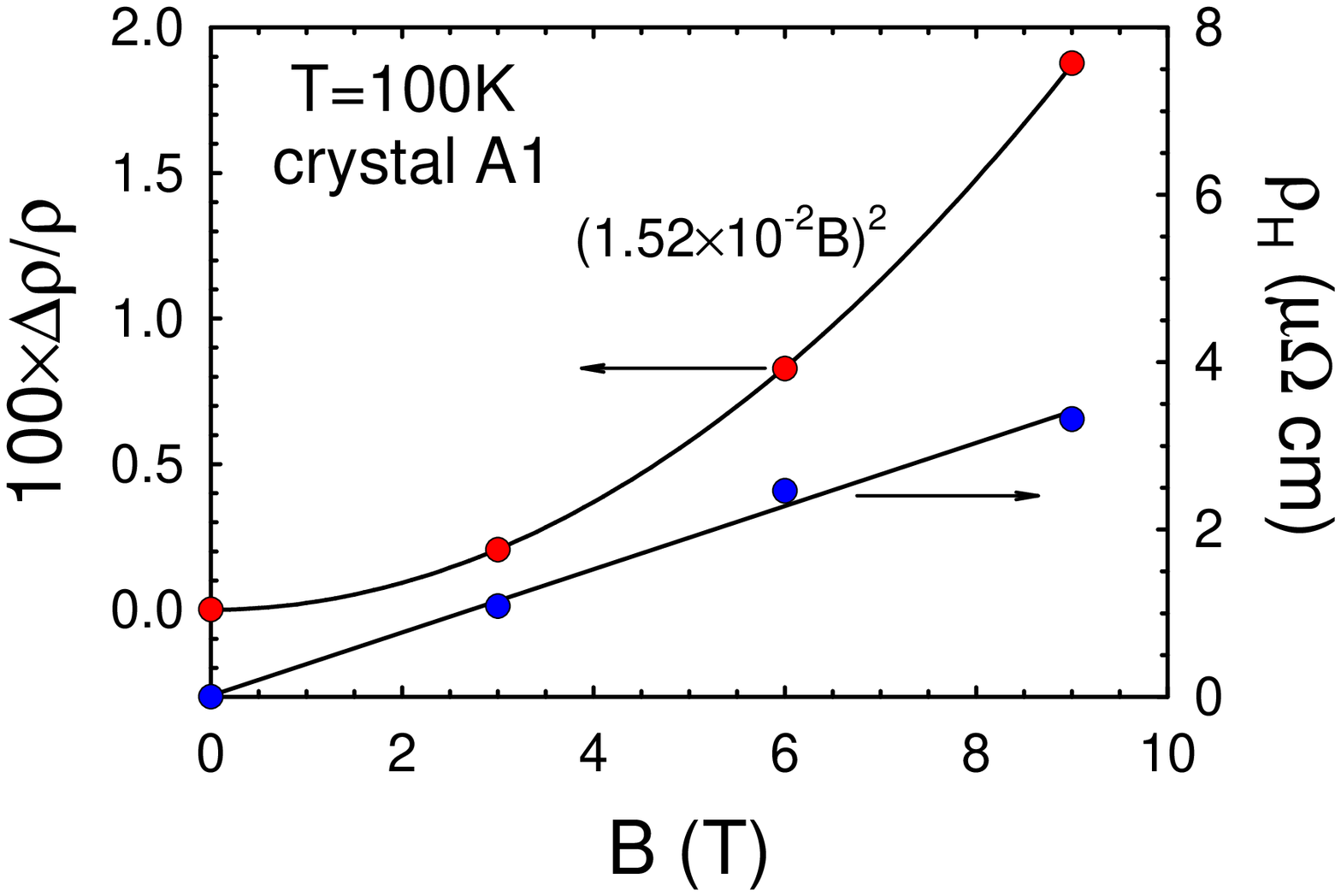}
\vglue -.2in
\caption{Magnetoresistance and Hall resistivity versus magnetic field for current applied along the conducting chains ($b$ axis)
and field along $a$.}
\end{figure}

At $T>50$~K the MR was quadratic and the Hall resistivity linear
in field (Fig.~S3), thus $R_{H\infty}$ is irrelevant. Simultaneously fitting the data at each temperature,
with the additional condition that the measured zero-field conductivity be equal to $\sigma^h+\sigma^e$, highly constrains the
four transport parameters.  The reliability of parameter sets (determined using the Levenberg-Marquardt method)
was established by varying the initial values of the parameters over a broad range.
The results for crystals A2 and C are shown in Fig.~S4.

\begin{figure}[t]
\includegraphics[width=5.4in,clip]{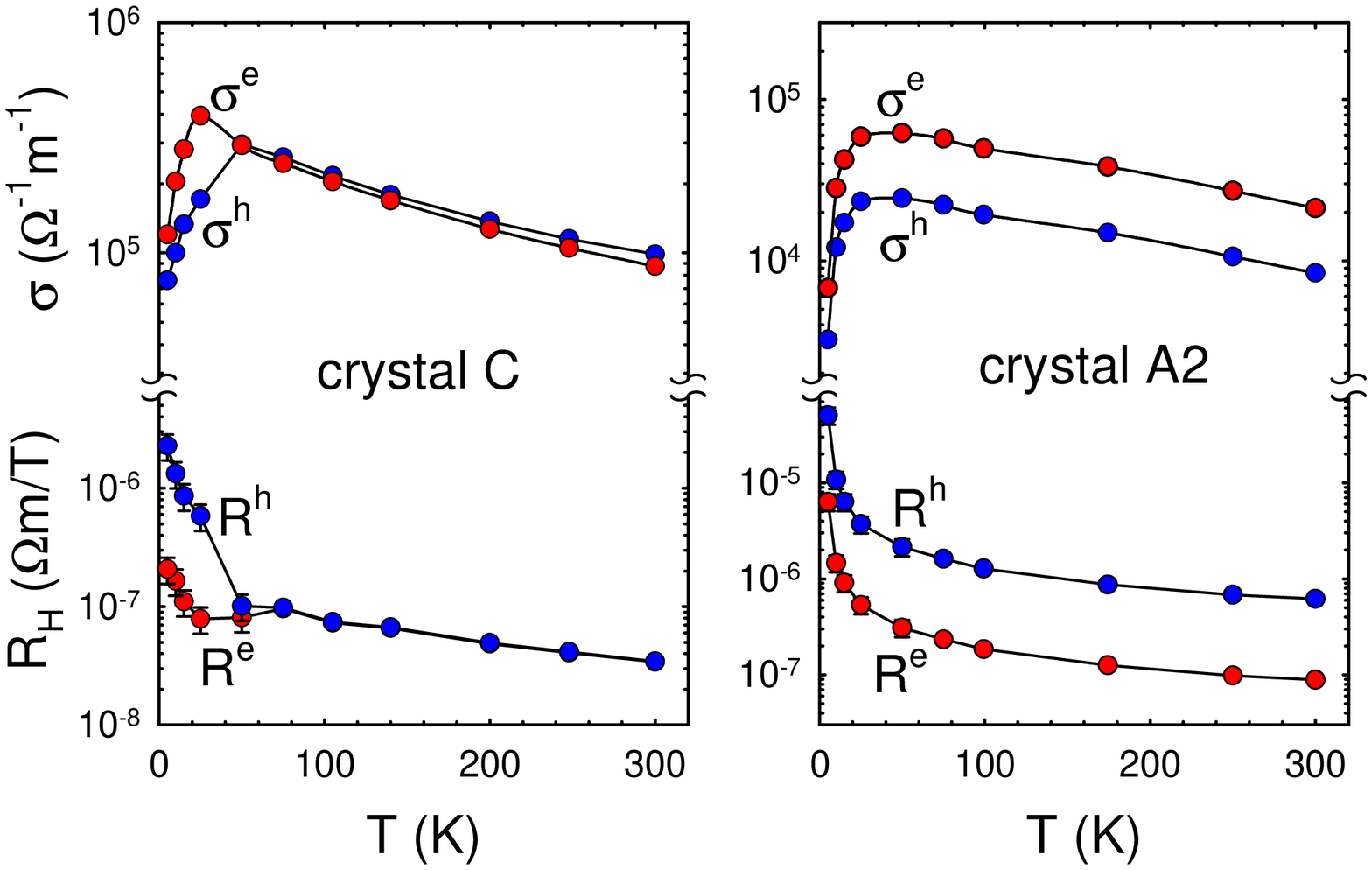}
\caption{Hall coefficients and conductivities for holes and
electrons along the conducting chains ($b$ axis) from two-band
model fitting. Error bars at $T\leq 50$~K reflect uncertainties
associated with deviations of the MR from the two-band form (noted
previously in Ref.~14).} \label{S4}
\end{figure}

\begin{figure}
\includegraphics[width=3.5in,clip]{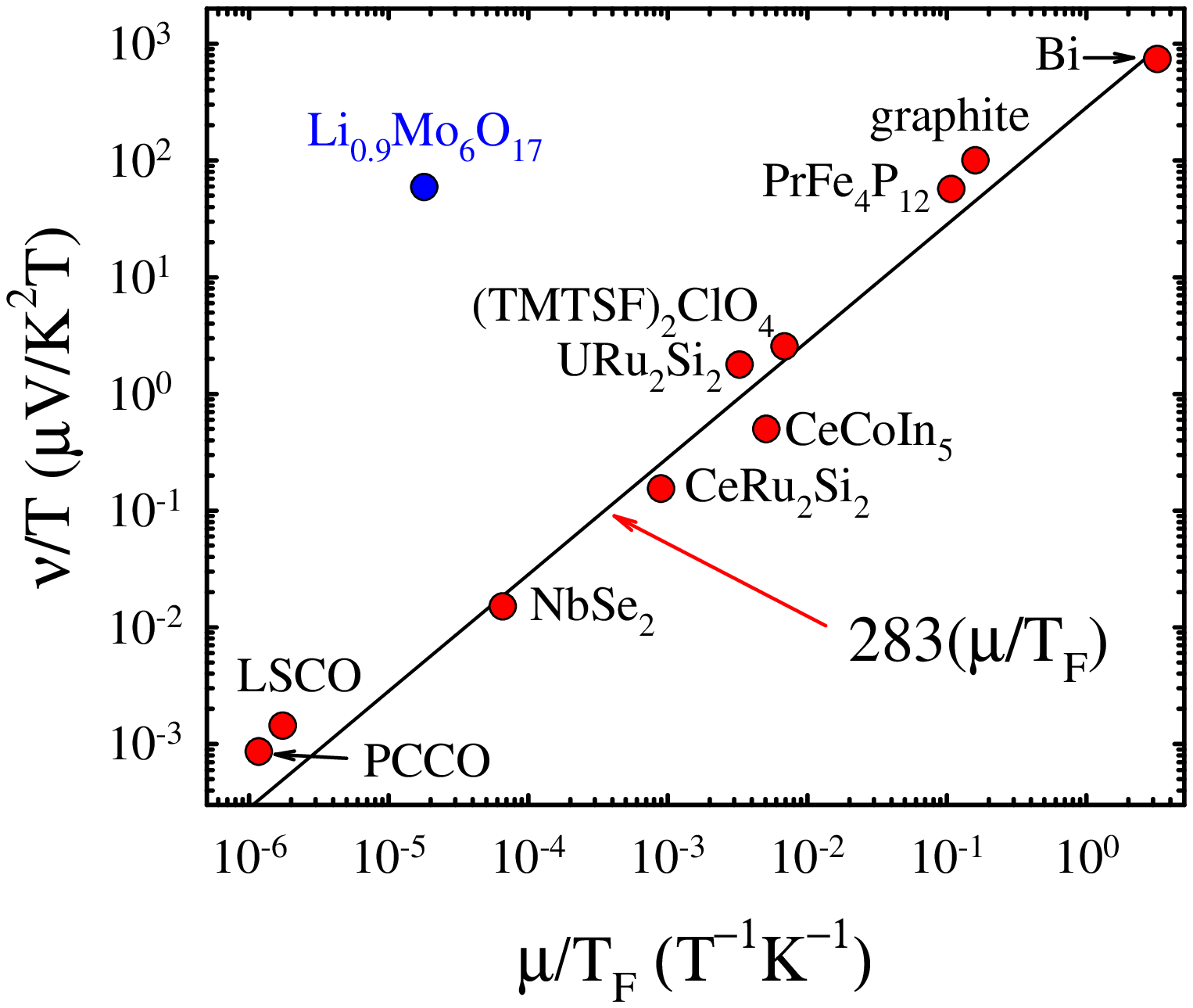}
\vglue -.7in
\caption{Low-$T$ $\nu/T$ plotted versus $\mu/T_F$ for specimen C and a variety of other compounds (adapted from Ref.~5).}
\label{S5}
\end{figure}
\vspace{-.2in}
\section*{Fermi temperature and mobilities}

The Fermi temperature can be estimated from the energy difference
between the Fermi level and bottoms of the two bands dispersing
along the $\Gamma$-Y direction of the Brillouin zone, as
determined by photoemission (0.25 and 0.49 eV, Ref.~8) or band
structure calculations (0.38 eV and 0.62 eV, Ref.~9),
i.e. $T_F\simeq (4-6)\times 10^3$~K. The dash-dot-dot curve
labeled ``$283\mu/T_F$'' in Fig.~1~a uses $T_F=5\times 10^3$~K and
$\mu=\mu^h+\mu^e$ computed as $\mu^i=\sigma^iR^i$~($i=h,e$) from
Fig.~S4.

\section*{Estimate of Fermi temperature from the thermopower}

Changes in the electronic structure at $T<T_m$ lead to a divergence of the TEP (Fig.~3) and thus preclude an estimate
of $T_F$ from the $T\to 0$ behavior of the diffusion thermopower.
Instead we estimate $T_F$ from the slopes of the linear-$T$ thermopowers for specimens C, D, and E, under the assumption they
reflect the diffusion contribution (we ignore the positive offsets in the TEP extrapolated to $T=0$ for these specimens
that are reminiscent of the TEP in cuprates \cite{Obertelli}).
The average value, $dS/dT=-0.045\mu{\rm V/K}^2$, combined with the Mott expression under
the assumption of a linear energy dependence of the conductivity at the Fermi level yields,
$T_F=(\pi^2/3)(k_B/e)(-dS/dT)^{-1}\simeq 6.3\times 10^3$~K.
This agrees satisfactorily with the $T_F$ estimates above given that
the linear slope of our data should be smaller than those of either electrons or holes due to partial cancellation,
$S=(\sigma^hS^h+\sigma^eS^e)/(\sigma^h+\sigma^e)$.

\section*{Carrier density estimated from chemistry/bonding}

The two $xy$ bands at $E_F$ in the unit cell of Li$_{0.9}$Mo$_6$O$_{17}$ (composed of two formula units)
are half full (Ref.~9), i.e. 1 electron per band or 2 mobile electrons per unit cell.  The Li deficiency depletes
charge (0.2 electrons per cell) from the chain Mo atoms, only half of which
contribute to occupied orbitals at $E_F$.  Thus the mobile charge per unit cell is estimated as $2-0.2/2=1.9$ electrons.
With the unit cell volume, $V=668$~\AA $^3$, this yields $n=2.84\times 10^{27}$~m$^{-3}$.  Oxygen deficiency should increase $n$ from this value.

\section*{Seebeck coefficients computed from anisotropic Nernst expression}

As noted in the text (Ref.~27), the Nernst coefficient for an
anisotropic, bipolar conductor can be written as,
\begin{equation}
\nu_{yx}=\left(\sigma_x\over\sigma_y\right){\sigma_x^h\nu_{yx}^h+\sigma_x^e\nu_{yx}^e\over
\sigma_{x}}+{(R^h\sigma_{y}^h-R^e\sigma_{y}^e)
\sigma_{x}^h\sigma_{x}^e(S_{x}^h-S_{x}^e)\over\sigma_{x}\sigma_{y}}
\tag{S1}
\end{equation}
\noindent where $R^h (R^e)$ and $S^h (S^e)$ are the hole (electron) Hall and partial Seebeck coefficients, respectively, and $\sigma_{x}\equiv \sigma_{x}^h+\sigma_{x}^e$, etc. for $\sigma_{y}$.
The Seebeck coefficient is defined as, $S_x=(\sigma_x^hS_x^h+\sigma_x^eS_x^e)/\sigma_{x}$.
For perfect compensation, $\sigma_{x}^h=\sigma_{x}^e$, $\sigma_{y}^h=\sigma_{y}^e$, $R^h=-R^e$, the second term reduces
to, $(1/2)\mu_x(S_x^h-S_x^e)$.  Note that $\sigma_{x}^h=\sigma_{x}^e$ does not necessarily imply $\sigma_{y}^h=\sigma_{y}^e$.
Since we have not measured the magnetoresistance for electric current along the \emph{c} axis,
for the calculations of $S_b^h$ and $S_b^e$ using eq.~(S1) described in the text, we took $\sigma_{c}^h/\sigma_{c}=\gamma (\sigma_{b}^h/\sigma_{b})$
with $\gamma$ as an adjustable parameter.  In addition to the case $\gamma=1$ (shown in the inset of Fig.~3) we also show in Fig.~S6
cases where holes are dominant for \emph{c}-axis conduction ($\gamma=1.8,\ 3$ for specimens A2, C) or electrons are dominant ($\gamma=0.1$).
Colossal Seebeck coefficients are found in either case.
\newpage
\vglue -.25in
\begin{figure}
\includegraphics[width=4.6in,clip]{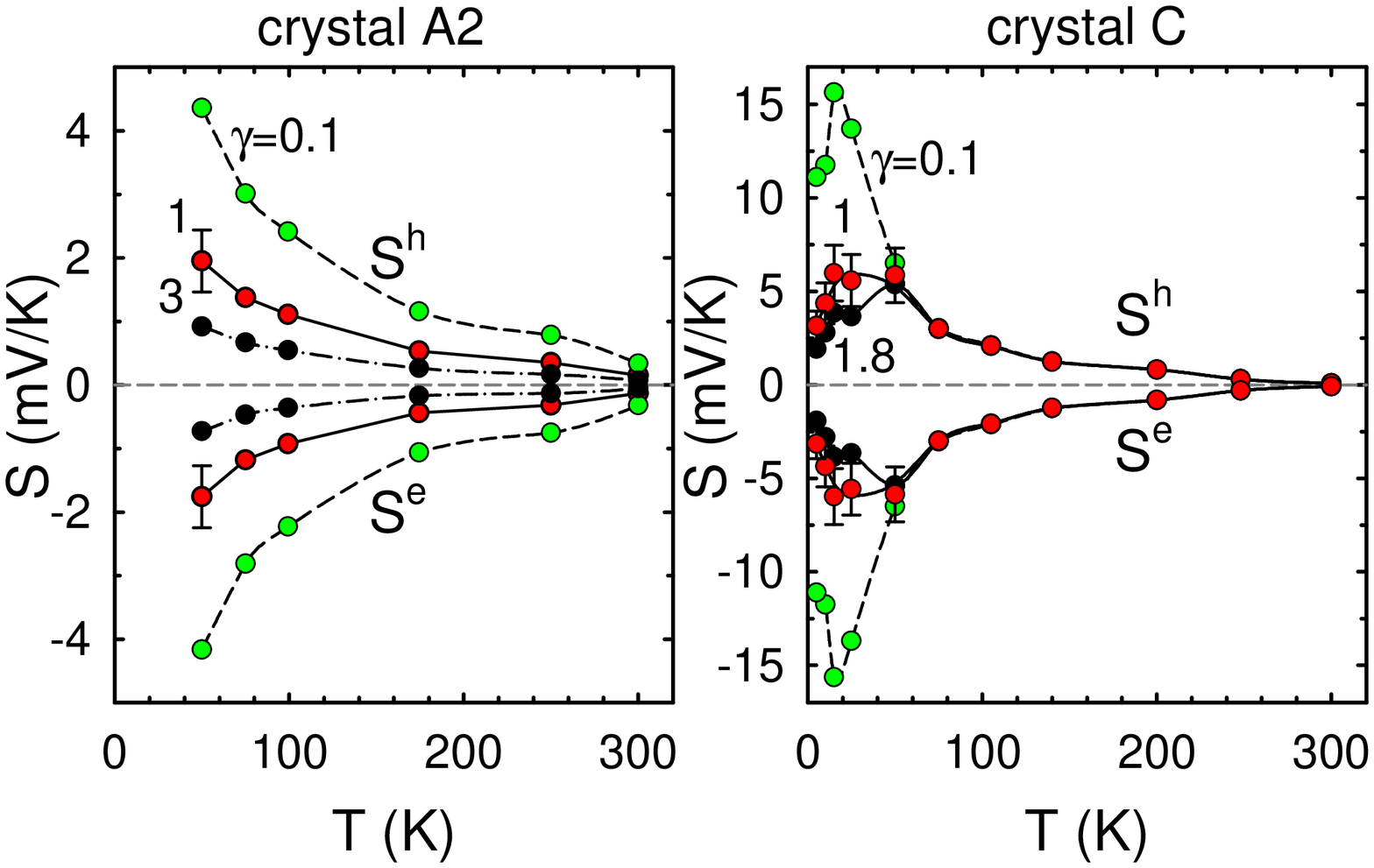}
\vglue -.15in
\caption{Hole and electron partial thermopowers computed using equation~(S1) and the two-band thermopower expression, with the measured
Nernst coefficients and two-band fit parameters from Fig.~S4 as input.  Error bars (25\%) are propagated from Fig.~S4.}
\label{S6}
\end{figure}
\begin{figure}
\includegraphics[width=3.5in,clip]{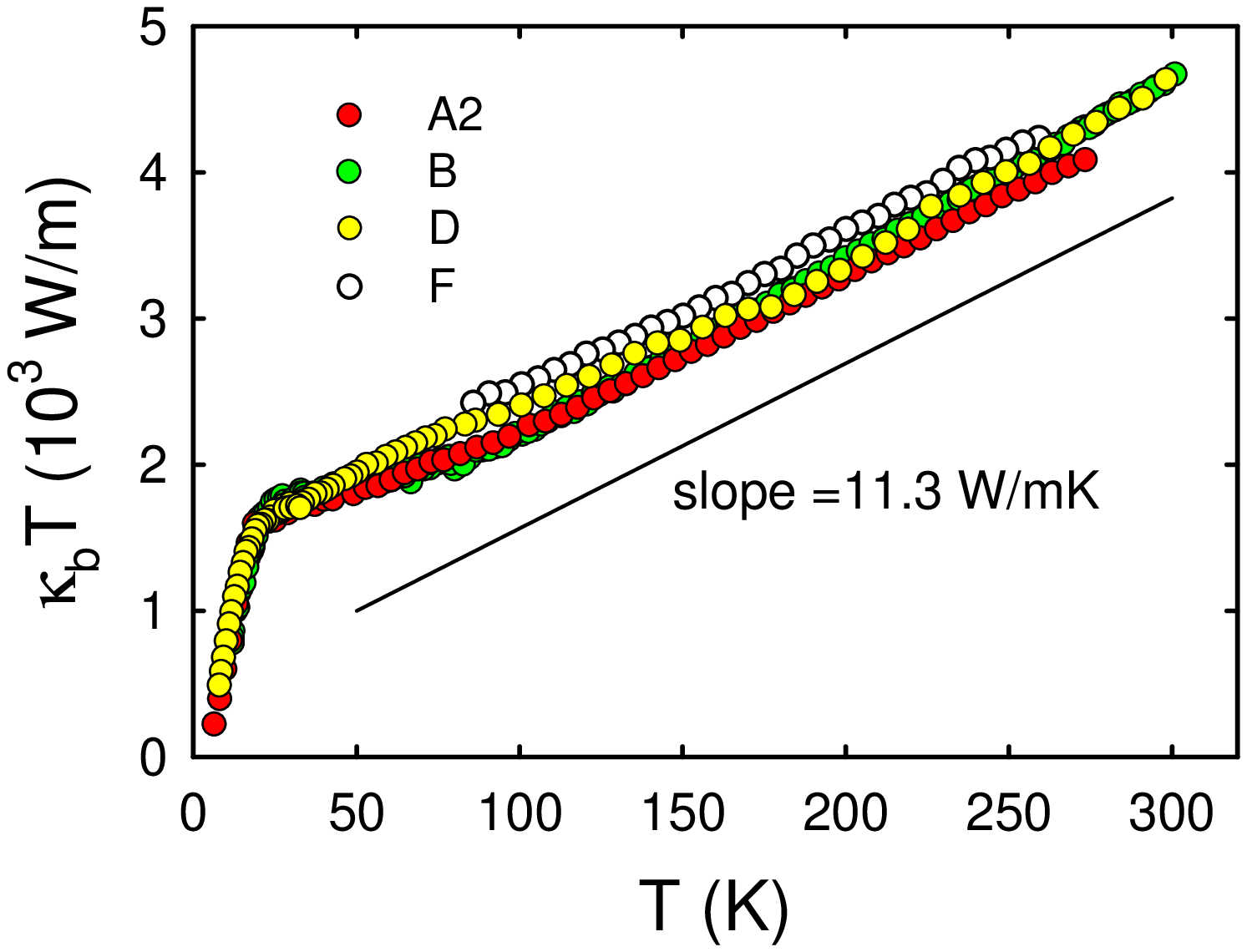}
\vglue -.15in
\caption{\emph{b}-axis thermal conductivites, plotted as $\kappa_bT$ \emph{vs.} $T$ for four other crystals with
$\sigma_b (300~{\rm K})$ ranging from 0.8-3 ${\rm m}\Omega {\rm cm}$. Crystal F, not listed in Table I, had a \emph{c}-axis lattice
constant and thermopower very similar to that of specimen D.}
\label{S7}
\end{figure}
\vglue -.2in

\end{document}